\documentclass[epj,nopacs]{svjour}
\usepackage[T1]{fontenc}
\usepackage[latin9]{inputenc}
\usepackage{color}
\usepackage{array}
\usepackage{amstext}
\usepackage{graphicx}
\usepackage{esint}
\usepackage{graphics}
\begin{document}
\title{Heavy quark perturbative QCD fragmentation functions
 in the presence of hadron mass}
\author{S. Mohammad Moosavi Nejad\inst{1,2}, Aida Armat\inst{1}}

\institute{Faculty of Physics, Yazd University, P.O. Box
89195-741, Yazd, Iran \and
School of Particles and Accelerators,
Institute for Research in Fundamental\\
Sciences (IPM), P.O.Box
19395-5531, Tehran, Iran}

\date{Received: date / Revised version: date}
%
\abstract{The dominant  mechanism to produce  hadronic bound states  with large transverse momentum
is fragmentation, that is the splitting of a high energy parton into a hadronic state and other partons.
We review the present schemes to calculate the heavy quark fragmentation functions (FFs)  and drive
an exact analytical expression of FF which includes most of the kinematical and dynamical
properties of the process. Using the perturbative QCD,  we calculate the FF for
 $c$-quark to split into $S$-wave $D^+$ meson to leading order in the QCD coupling constant.
Our result is compared with  the current well-known phenomenological models which  are obtained through a global fit
to $e^+e^-$ data from SLAC SLC and CERN LEP1 and  we also compare the FF
with experimental data form BELLE and CLEO.  Specifically, we study the effect of outgoing meson mass
on the pQCD FF. Meson masses are responsible for the low-$z$ threshold, where $z$ is the scaled energy variable.}

\maketitle

\section{Introduction}
\label{sec:intro}

Hadron production processes are important  in  investigating properties of quarks in heavy ion collisions and in finding
the origin of the nucleon spin in lepton-nucleon scattering  processes and polarized proton-proton
collisions. In order to calculate
the hadron production cross section, the fragmentation functions (FFs) are the key quantities
and they must be known in advance.  The FFs describe hadron production probabilities from the initial partons
and they cannot be precisely calculated by theoretical approaches  at this stage.
The FFs are related to the low energy part  of the hadron production processes
and they form the nonperturbative aspect of QCD.  The FFs are universal and
their importance is for model independent predictions of the cross
sections at the Large Hadron Collider (LHC) in which a hadron is detected in the outgoing
productions as a colorless bound state. They can also be used to find the internal structure of the
exotic hadrons using the differences between the disfavored and favored FFs \cite{Hirai:2010cs}.
The QCD improved parton model provides a great  theoretical frame to extract these functions.
However, once they
are given at the initial fragmentation scale $\mu_0$, their $\mu$ evolution is determined by
the Dokshitzer-Gribov-Lipatov-Alteralli-Parisi (DGLAP)
 renormalization group equations \cite{dglap} which are very similar to
 those for parton distribution functions  (PDFs).
The universality of the initial condition of the FFs, first was suggested  in \cite{Mele}
in the framework of  $e^-e^+$ annihilation and afterward  was proved in a more
general way in Ref.~\cite{Cacciari:2001cw}.\\
There are two main approaches to evaluate the FFs.
In the first approach, which is frequently used to obtain
the FFs, these functions are extracted from experimental data analyses instead of theoretical calculations.
 In this scheme, which is normally called the phenomenological approach,
the FFs are mainly determined by hadron production data
of $e^-e^+$ annihilation, lepton-hadron and hadron-hadron scattering 
processes by working either in $x$-space \cite{fragfunction,Albino:2005me,Kniehl:2000fe,Kneesch:2007ey}
or in Mellin-N space  \cite{Nason:1999zj,Cacciari:2005uk}.
This situation is very similar to the determination of the PDFs.
In this approach, according to the Collin's factorization theorem \cite{Collins:1998rz}
 the cross section of hadron production in the $e^-e^+$ annihilation is described by the convolution
of  partonic hard-scattering cross sections ($e^-e^+\rightarrow q\bar q$) which
are calculable perturbatively and a
realistic fragmentation function describing the transition
of a parton into an outgoing hadron. In this scheme, the FFs involve parameters to be fixed by  fitting the experimental data.
Various phenomenological models like Peterson model \cite{Peterson:1982ak}, Lund model \cite{Andersson:1983ia},
Cascade model \cite{Webber:1983if} and etc, have been developed to describe the FFs.

The second approach is based on this fact that  the FFs for mesons containing a heavy quark can be computed theoretically using
perturbative QCD (pQCD) \cite{Ma:1997yq,Braaten:1993rw,Chang:1991bp,Braaten:1993mp,Scott:1978nz}.
The first theoretical attempt to explain the procedure of hadron production by a heavy quark was made by Bjorken \cite{Bjorken:1977md} by using a
naive quark-parton model (QPM). He deduced that the inclusive distribution of  heavy hadron  should peak
almost at $z=1$, where $z$ refers to  the scaled energy variable. This property
is important for heavy quarks for which the peak of heavy quark fragmentation function occurs closer to $z=1$.
In continuation, Peterson  \cite{Peterson:1982ak} presented the popular form of FF
which manifestly behaves as $(1-z)^2$ at large $z$ values, using a quantum mechanical parton model.
The pQCD scheme was followed by Suzuki \cite{Suzuki:1977km},  Amiri and Ji \cite{Amiri:1986zv}. While
in this scheme Suzuki calculates the heavy FFs using a diagram similar to that in Fig.~\ref{ff}.\\
Here, we focus on heavy quark FFs and drive an exact analytical
form of FF, using the Suzuki's approach which embeds most of the kinematical and
dynamical properties of the process. Our results are compared both with one of the well-known
phenomenological models and with the experimental data and they are found  in good agreement.
Furthermore, we impose finite meson mass effect on the perturbative QCD FF. This modifies the relations between partonic and
hadronic variables and reduces the available phase space and 
is responsible for the low-z threshold. 
In Ref.~\cite{Kneesch:2007ey}, the effect of finite hadron mass
on the non-perturbative fragmentation function is studied and it is shown that
the inclusion of finite mass effect tends to improve the overall description of
the data. Specifically, it is shown that hadron mass effect turned out
to be more important than quark mass effects.
Although this additional effect is
not expected to be truly sizable numerically, its
study is nevertheless necessary in order to fully exploit the enormous statistics of the LHC data
to be taken in the long run for a high-precision determination of the top-quark properties.

This paper is organized as follows.
In Sec.~\ref{sec:one}, we explain the phenomenological approach to calculate the FFs by
introducing a well-known  model.
In Sec.~\ref{sec:two}, the theoretical scheme to calculate the FFs is introduced in detail.
We then discuss the use of the pQCD fragmentation functions as a phenomenological model
for the fragmentation of the charm quark into the heavy-light mesons $D^0$ and $D^+$.
In Sec.~\ref{sec:three} we study, for the first time, the effect of meson mass on the perturbative 
QCD FFs and we present the numerical results
and in Sec.~\ref{sec:four},   our conclusions are summarized.

\section{Determination of fragmentation functions: Phenomenological scheme}
\label{sec:one}

One of the most current approaches to determine the FFs is the method
based on  data analyzing. The FFs are studied in hadron-hadron, lepton-hadron deep inelastic
scattering (DIS) and electron-positron annihilation.
Among  all, the FFs are mainly determined by hadron production data
of $e^-e^+$ annihilation.  The perturbative QCD framework is used to study single-inclusive hadron production
in $e^-e^+$ annihilation, where the factorization theorem is an important tool to study this process.
According to the factorization theorem of the QCD improved parton model \cite{Collins:1998rz},
 in the high energy scattering  the cross section can be expressed in terms of the partonic hard scattering cross
 sections and the non-perturbative FFs $D_i^H(z, Q^2)$ in which the last one is related to
the low energy components of process, i.e.
\begin{eqnarray}\label{fac}
\frac{1}{\sigma_{tot}}\frac{d}{dz}\sigma(e^+e^-\rightarrow HX)=\sum_i C_i(z, \alpha_s) \otimes D_i^H(z, Q^2),\nonumber\\
\end{eqnarray}
where, the function $D_i^H(z, Q^2)$ indicates the probability to find the
hadron $H$ from a parton $i(=g, u, d, s, \cdots)$ with the energy fraction $z$
and $C_i(z, \alpha_s) $ is the Wilson coefficient function based on the partonic cross section $e^+e^-\rightarrow q\bar q$
which is calculated in the perturbative QCD \cite{Kniehl:2000fe,Nason:1993xx}, and the convolution
integral is defined as $f(z)\otimes g(z)=\int_z^1 dy/y f(y)g(z/y)$. In the equation above,
 $X$ stands for the unobserved jets and  $\sigma_{tot}$  is the total hadronic
cross section \cite{Kneesch:2007ey}, and $Q^2$ is the squared center-of-mass energy
$s=Q^2$. The variable $z$ stands for the fragmentation parameter and is defined by the energy fraction
\begin{eqnarray}\label{par1}
z=\frac{2E_H}{\sqrt{s}},
\end{eqnarray}
where  $E_H$ is the  energy of detected hadron. In fact, the fragmentation parameter refers  to the energy fraction of process
which is taken away by the outgoing hadron $H$.\\
In the phenomenological approach, the FFs are parameterized in a convenient functional form at  the initial
scale $\mu_0^2$ in each order, i.e. LO and NLO. The initial scale $\mu_0^2$ is different for partons and the initial FFs
 are evolved to the experimental $\mu^2$ points by
the DGLAP evolution equations \cite{dglap}.
The FFs are parameterized in terms of a number of free parameters which are determined by an $\chi^2$ analysis of the
$e^+e^-\rightarrow H+X$ data at the scale $\mu^2=s$.  Due to the energy conservation, there is the following  constraint
for the parameters
\begin{eqnarray}
\sum_H \int_0^1 dz z D_i^H(z, Q^2)=1.
\end{eqnarray}
This constraint is known as the energy sum rule, which means that each parton will fragment into some hadrons $H$.\\
In Ref.~\cite{Kniehl:2011bk}, authors calculated the $b\rightarrow B$ FFs based on the Peterson and power ansaetze
obtained through a global fit to $e^+e^-$ data from CERN LEP1 and SLAC SLC. In Ref.~\cite{Kneesch:2007ey}, authors
reported the FFs for $D^0, D^+$ and $D^{\star +}$ mesons by fitting the experimental data from
the BELLE, CLEO, ALEPH, and OPAL collaborations in the modified minimal-subtraction ($\overline{MS}$) factorization scheme.
They have parameterized the $z$ distributions of the $c$ and $b$ quark FFs at their starting scales $\mu_0=m_c$ and $m_b$, respectively, as
suggested by Bowler \cite{Bowler}, as
\begin{eqnarray}\label{bw}
D_q^{H_c}(z,\mu_0)=Nz^{-(1+\gamma^2)}(1-z)^a e^{-\gamma^2/z},
\end{eqnarray}
with three free parameters.  As they claimed, this parametrization yields the best fit to the BELLE
data \cite{Seuster:2005tr} in a comparative analysis using the Monte Carlo event generator JETSET/PYTHIA.
The values of fit parameters for the $D^+(c\bar d)$ meson are obtained from the BELLE/CLEO, OPAL, and the
global fits   using  the massless scheme or  zero-mass variable-flavor-number (ZM-VFN) scheme \cite{jm} where
$m_q=0$ is put from the beginning and the non-zero values of the  $c$ and $b$ quark masses only enter through the initial
conditions of the non-perturbative FFs.
The values of fit parameters
together with the  achieved values of $\overline{\chi^2}$ are reported in Table \ref{fit}.

\begin{table}[t]
\caption{\label{fit} Values of fit parameters for $c\rightarrow D^+$ FF
at the starting scale $\mu_0=m_c=1.5$ GeV obtained
from the Belle/CLEO, OPAL, and global fits in the ZM
approach together with the   values of $\overline{\chi^2}$ achieved.}

\begin{tabular}{ccccc}
\hline
 &$N$ & $a$ & $\gamma$ &$\overline{\chi^2}$ \\
\hline
Belle/CLEO-ZM &$ 7.30\times 10^5$&$1.12$& $3.43$&$1.37$
\\
OPAL-ZM & $2.62\times 10^4$&$1.48$&$2.91$&$0.507$
\\
Global-ZM &$7.31\times 10^5$&$1.13$& $3.43$&$2.21$
\end{tabular}
\end{table}

\section{Heavy quark fragmentation functions: Perturbative QCD  scheme}
\label{sec:two}

\begin{figure}
\begin{center}
\includegraphics[width=1.4\linewidth,bb=99 562 392 729]{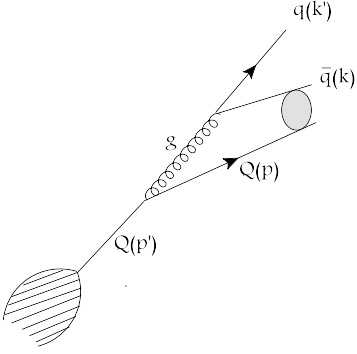}
\caption{\label{ff}%
Formation of a heavy meson. A heavy quark $Q$ forms a bound state $Q\bar q$
with a light antiquark produced through a single vector gluon.}
\end{center}
\end{figure}

As  is pointed out in Refs.~\cite{Braaten:1993rw,Chang:1991bp,Braaten:1993mp}
the fragmentation function $D_{Q\rightarrow M}(z,\mu_0)$ for meson $M$ containing a heavy quark
or a heavy antiquark $Q$ can be computed using the perturbative QCD.
Here, the fragmentation parameter $z$ is
the longitudinal momentum fraction of hadron relative to the fragmenting heavy
quark, i.e.
\begin{eqnarray}\label{par2}
z=\frac{(E+p_{||})_{hadron}}{(E+p_{||})_{beam}}.
\end{eqnarray}

\begin{figure}
\begin{center}
\includegraphics[width=1\linewidth,bb=37 192 552 629]{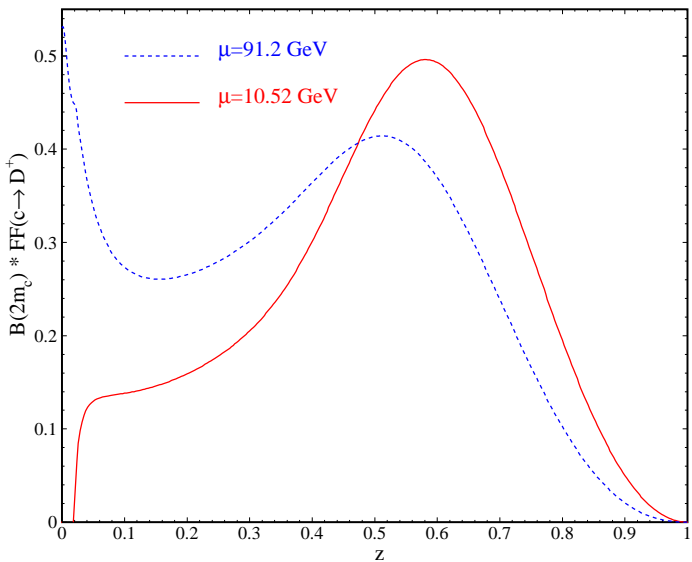}
\caption{\label{dglap}%
The fragmentation function $D_c^{D^+}$ as a function of $z$ for $\mu=10.52$ GeV (solid line) and  $\mu=m_z$ (dashed line)
 normalized by the branching fraction $B(2 m_c)$.}
\end{center}
\end{figure}

\begin{figure}
\begin{center}
\includegraphics[width=1.0\linewidth,bb=88 610 322 769]{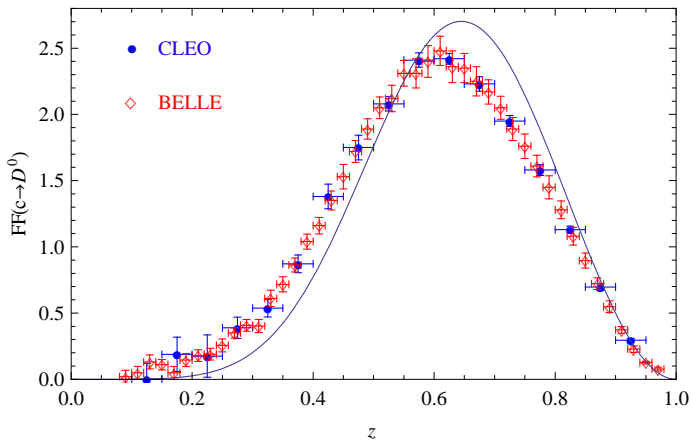}
\caption{\label{dzero}%
Comparision of pQCD  FF with data from BELLE and CLEO on $D^0$ production at the initial scale $\mu_0=2m_c$. 
We set $\left\langle k_\bot^2 \right\rangle=m_c^2$.}
\end{center}
\end{figure}

\begin{figure}
\begin{center}
\includegraphics[width=1.0\linewidth,bb=88 610 322 769]{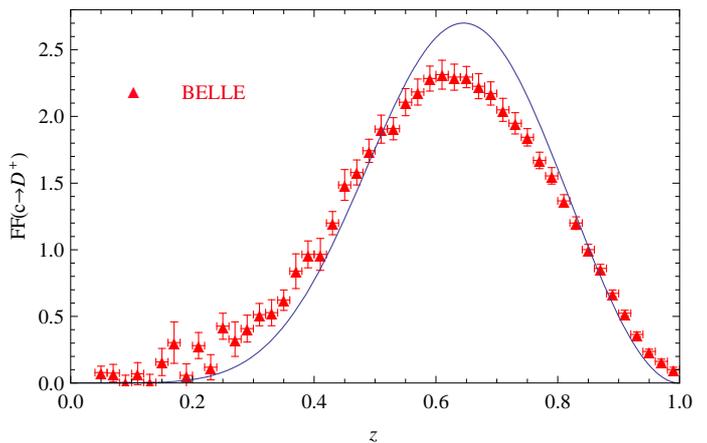}
\caption{\label{dplus}%
As in figure \ref{dzero}, but comparing with $D^+$ data from BELLE experiments at the initial scale $\mu_0=2m_c$.}
\end{center}
\end{figure}

In this work, using pQCD we apply the theoretical approach proposed by Suzuki \cite{Suzuki:1985up} which is independent of
data analyzing and is based on the convenient
Feynman diagrams and the wave function of the heavy meson bound state.
Therefore, at first, we briefly explain the Suzuki's  approach to obtain the analytical FF of a heavy quark
$Q$ into a heavy meson $M$ with  a bound state $Q\bar q$.
The main Feynman diagram for $Q\rightarrow M(Q\bar q)+q$ in the order of $\alpha_s^2$
including the four-momenta  is shown in Fig.~\ref{ff}.
According to the Lepage-Brodsky's approach \cite{Lepage:1980fj}, by neglecting the relative motion of  $Q$ and $\bar q$, we
 assume for simplicity that  $Q$ and $\bar q$ are emitted
collinearly with each other and they  move  along the $z$-axes.
Following Ref.~\cite{Suzuki:1985up}, we also adopt the infinite momentum frame where
 the fragmentation  parameter $z$ (\ref{par2}) is reduced to another popular form as
\begin{eqnarray}\label{par3}
z=\frac{E_{meson}}{E_{beam}}.
\end{eqnarray}
We also set the relevant four-momenta in Fig.~\ref{ff} as
\begin{eqnarray}\label{kinematic}
p_\mu^\prime &=&[p_0^\prime, \vec{k_\bot}, p_L^\prime] \quad   p_\mu=[p_0, \vec{0}, p_L] \nonumber\\
k_\mu^\prime &=&[k_0^\prime, \vec{k_\bot}, k_L^\prime]  \quad     k_\mu=[k_0, \vec{0}, k_L],
\end{eqnarray}
and the  momentum of the produced meson is set as $\bar{p}_\mu=[\bar{p}_0, \vec{0}, \bar{p}_L]$ where $ \bar{p}_L=p_L+k_L$.
We also may write the quark energies in terms of the  initial heavy quark energy $p_0^\prime$ and the fragmentation parameter as
\begin{eqnarray}
p_0=x_1 z p_0^\prime, \quad k_0=x_2 z p_0^\prime, \quad k_0^\prime=(1-z) p_0^\prime,
\end{eqnarray}
where $x_1=p_0/\bar{p}_0$ and $x_2=k_0/\bar{p}_0$ are  the meson energy fractions carried by the constituent quarks.
As  in Ref.~\cite{GomshiNobary:1994eq}, it is also assumed  that the contribution of each constituent quark from the meson energy is proportional
to its mass, i.e. $x_1=m_Q/M$ and $x_2=m_{\bar q}/M$ where $M=m_Q+m_{\bar q}$.\\
Following Refs.~\cite{Suzuki:1985up,GomshiNobary:1994eq}, the fragmentation function may be defined as
\begin{eqnarray}\label{first}
D_{Q\rightarrow M}(z, \mu_0)&=&\frac{1}{\sigma}\frac{d\sigma}{dz}\nonumber\\
&&\hspace{-1.5cm}=\int d^3\vec{p} d^3\vec{k} d^3\vec{k^\prime} \overline{|T_{Q\rightarrow M}|^2} \delta^3(\vec{k}+\vec{p}+
\vec{k^\prime}-\vec{p^\prime}),\nonumber\\
\end{eqnarray}
where the average probability amplitude squared $\overline{|T|^2}$ is obtained as $\sum_{s}|T|^2 /(1+2s_Q)$ in which  the summation is going
over the spins and colors  and   $s_Q$ is the initial heavy quark spin.  The probability amplitude $T_{Q\rightarrow M}$
 is expressed as the convolution of the hard scattering amplitude $T_H$, which can be computed perturbatively from quark-gluon
 subprocesses, and the process-independent distribution amplitude $\Phi_M$ which contains 
 the bound state nonperturbative dynamic of outgoing meson, i.e.
\begin{eqnarray}\label{base}
T_{Q\rightarrow M}=\int dx_1dx_2\delta(1-x_1-x_2) T_H(x_i,Q^2) \Phi_M(x_i, Q^2).\nonumber\\
\end{eqnarray}
This scheme,  introduced in \cite{Adamov:1997yk,brodsky}, is used to absorb the soft behavior of the bound state into the  hard scattering amplitude $T_H$.
In (\ref{base}),  $\Phi_M$ is the probability amplitude to find the quarks which are collinear in
 the mesonic bound state up to the scale $Q$.
 In general, the probability amplitude $\Phi_M$ is related to the hadronic wave function $\Psi_M$ by
 \begin{eqnarray}\label{formul}
\Phi_M(x_i, Q^2)=\int [d^2 \vec{q_{\bot i}}]\Psi_M(x_i, \vec{q_{\bot i}})\Theta( \vec{q_{\bot i}}^2<Q^2),\nonumber\\
\end{eqnarray}
where
\begin{eqnarray}
 [d^2 \vec{q_{\bot i}}]=2 (2\pi)^3\delta\big[\sum_{j=1}^2  \vec{q_{\bot j}}\big]\prod_{i=1}^2\frac{d^2 \vec{q_{\bot i}}}{2(2\pi)^3}.
\end{eqnarray}
The probability amplitude $\Phi_M$ represents the valence quark and antiquark wave function evaluated at quark impact separation $ b_\bot\approx Q^{-1}$.
 Here, $\Theta(x)=\int_{-\infty}^x dt \delta(t)$ is the Heaviside step function 
 and  $ \vec{q_{\bot i}}$ refers to the  transverse momentum of constituent quarks.  A typical simple mesonic wave function is
  \begin{eqnarray}\label{formula}
\Psi_M(x_i, \vec{q_{\bot i}})=\frac{(128 \pi^3 b^5 M)^{\frac{1}{2}}}{x_1^2 x_2^2\big[M^2-\frac{m_1^2+\vec{q_{\bot 1}}^2}{x_1}-
\frac{m_2^2+\vec{q_{\bot 2}}^2}{x_2}\big]^2},\nonumber\\
\end{eqnarray} 
 where $M$ is the meson mass and $b$ is the binding energy of the two body bound state.
 Both in the case $m_1=m_2$ and $m_1>>m_2$, it can be shown that the above wave function is the solution of the
 Schr\"{o}dinger equation with a Coulomb potential, which is the nonrelativistic limit of the Bethe-Salpeter equation with the QCD kernel \cite{brodsky}.\\
Working in the infinite-momentum frame and by considering  (\ref{formul}) and (\ref{formula}) we integrate 
over $\vec{q_{\bot i}} (0\leq\vec{q_{\bot i}}^2\leq \infty)$ where $\vec{q_{\bot i}}$ stands for $\vec{q_{\bot 1}}$ or $\vec{q_{\bot 2}}$ .
 The integration yields an expression as
 \begin{eqnarray}
\Phi_M(x_i, Q^2)=\frac{(128 \pi b^5 M)^{\frac{1}{2}}}{16\pi^2(x_1+x_2) (m_1^2 x_2+m_2^2 x_1-x_1 x_2 M^2)},\nonumber\\
\end{eqnarray} 
 which grows rapidly  at $x_1=1-x_2=m_1/M$ when $M$ is set to $m_1+m_2$ and therefore
 is estimated as a delta function \cite{Amiri:1985mm}. In conclusion 
for a S-wave pseudoscalar heavy meson ($ ^{1}S_{0}$) with neglecting the Fermi motion, 
the probability amplitude  at large $Q^2$ reads
\begin{eqnarray}
\Phi_M\approx\frac{f_M}{2\sqrt{3}} \delta(x_1-\frac{m_1}{m_1+m_2}),
\end{eqnarray}
where $f_M=(6b^3/\pi M)^\frac{1}{2}$ refers to the decay constant for the meson.
The delta-function form is convenient for our assumption where  the constituent quarks inside the meson will fly 
together in  parallel and they have no transverse momentum.\\
Considering Fig. \ref{ff}, in which we make a simple approximation to form a
meson by emitting only a single gluon, the hard scattering  amplitude $T_H$ is expressed as
\begin{eqnarray}\label{forth}
T_H=\frac{2\pi \alpha_s m_qM_Q}{\sqrt{2p_0 k_0 k_0^\prime p_0^\prime}}\frac{C_F}{(k+k^\prime)^2 D_0}
\{\bar{u}(p)\gamma^\mu u(p^\prime)\bar{u}(k^\prime)\gamma_{\mu}v(k)\},\nonumber\\
\end{eqnarray}
where $D_0=p_0+k_0+k_0^\prime-p_0^\prime$ is the energy
denominator, $C_F=(N_c^2-1)/(2N_c)=4/3$ for $N_c=3$ quark colors, $\alpha_s$ is
 the strong coupling constant and $1/(k+k^\prime)^2$ is due to the gluon propagator.
Note that since the initial heavy quark is not on its mass shell, we have no energy conservation and thus we have
performed the energy integration to reproduce the energy denominator $D_0$.\\
To obtain the FF for an unpolarized meson, considering  (\ref{first}-\ref{forth}) and performing
an average over the initial spin states and a sum over the final spin states we find
\begin{eqnarray}
D_{Q\rightarrow M}(z, \mu_0)&=&N\int \frac{d^3\vec{p} d^3\vec{k} d^3\vec{k^\prime} \delta^3(\vec{k}+\vec{p}+
\vec{k^\prime}-\vec{p^\prime})}{D_0^2(k+k^\prime)^4p_0k_0 k^\prime_0 p^\prime_0}\times\nonumber\\
&&\bigg[2m_q^2m_Q^2-m_q^2(p^\prime\cdot p)+m_Q^2(k\cdot k^\prime)\nonumber\\
&&-(p\cdot k)(k^\prime\cdot p^\prime)-
(k\cdot p^\prime)(p\cdot k^\prime)\bigg],
\end{eqnarray}
where $N$ is proportional to $(\pi C_F \alpha_s f_M)^2$ but it is related to the normalization condition
 $\int_0^1 D_{Q}^M(z, \mu_0) dz=1$  \cite{Amiri:1986zv,Suzuki:1985up}.\\
To do the phase space integrations we consider the following integral
\begin{eqnarray}
&&\int \frac{d^3\vec{p}\delta^3(\vec{k}+\vec{p}+
\vec{k^\prime}-\vec{p^\prime})}{p_0 D_0^2}\nonumber\\
&&=\int \frac{d^3\vec{p}\delta^3(\vec{k}+\vec{p}+
\vec{k^\prime}-\vec{p^\prime})}{p_0(p_0+k_0+k_0^\prime-p_0^\prime)^2}\nonumber\\
&&=\int \frac{p_0d^3\vec{p}\delta^3(\vec{k}+\vec{p}+
\vec{k^\prime}-\vec{p^\prime})}{[p_0^2+p_0(k_0+k_0^\prime-p_0^\prime)]^2}\nonumber\\
&&=\frac{p_0}{[m_Q^2+(\vec{p}^\prime-(\vec{k}+\vec{k}^\prime))^2-p_0(p_0^\prime-(k_0+k_0^\prime))]^2}\nonumber\\
&&=\frac{p_0}{[(k+k^\prime)^2]^2},
\end{eqnarray}
where considering (\ref{kinematic}) one has
\begin{eqnarray}
(k+k^\prime)^2=-m_qM J^{\frac{1}{4}}(z, k_\bot^2 ),
\end{eqnarray}
with
\begin{eqnarray}
J(z, k_\bot^2 )=[1-2\frac{m_q}{M}-\frac{1}{z}-\frac{m_q^2+ k_\bot^2}{M^2}(\frac{z}{1-z})]^4,
\end{eqnarray}
and for the remaining integral we have
\begin{eqnarray}
\int d^3\vec{k^\prime} f(z, \vec{k_\bot^2})&=&\int dk^{\prime}_L d^2k_\bot f(z, \vec{k_\bot^2})\nonumber\\
&\cong& m_q^2 k^{\prime}_0 f(z, \left\langle k_\bot^2 \right\rangle),
\end{eqnarray}
where, for simplicity, we have replaced the transverse momentum integration by its average value $ \left\langle k_\bot^2 \right\rangle$, 
which is a free parameter and can be determined experimentally.\\
Putting all in (\ref{first}) we obtain the fragmentation function as
\begin{eqnarray}\label{last}
D_{Q\rightarrow M}(z, \mu_0^2)&=&\frac{N}{z(1-z)J(z, \left\langle k_\bot^2 \right\rangle)}\bigg\{RH\frac{m_q}{M^2}\nonumber\\
&&+\frac{z}{M}\big[\frac{TR}{1-z}-TH-m_Qm_q R\big]+\nonumber\\
&&2m_Qz^2 \big[T+m_q^2(1-z)(4-\frac{H}{zMm_Q})\big]\bigg\},\nonumber\\
\end{eqnarray}
where,
\begin{eqnarray}
T(z,\left\langle k_\bot^2 \right\rangle)&=&m_q^2+m_Q^2(1-z)^2+z^2\left\langle k_\bot^2 \right\rangle,\nonumber\\
H(z,\left\langle k_\bot^2 \right\rangle)&=&M^2+z^2(m_Q^2+\left\langle k_\bot^2 \right\rangle),\nonumber\\
R(z,\left\langle k_\bot^2 \right\rangle)&=&M^2(1-z)^2+z^2(m_q^2+\left\langle k_\bot^2 \right\rangle).
\end{eqnarray}
\begin{figure}
\begin{center}
\includegraphics[width=1\linewidth,bb=37 192 552 629]{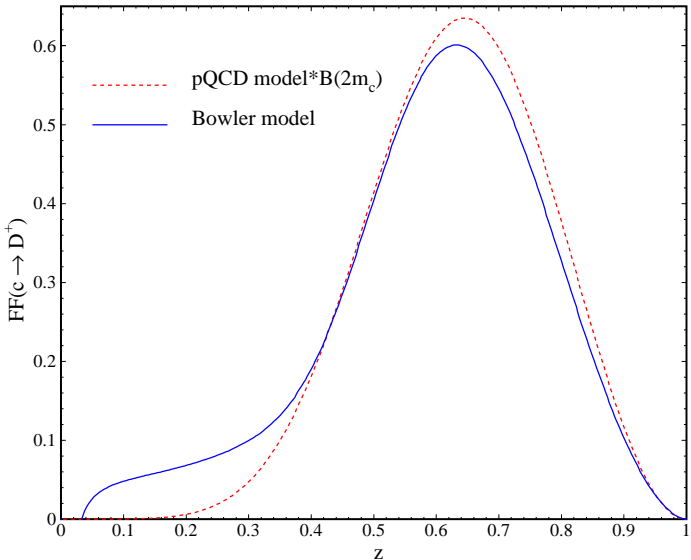}
\caption{\label{comparision}%
$c\rightarrow D^+$ FF at the initial scale $\mu_0=2 m_c$ as a function of $z$ in the pQCD approach (dashed line)
and Bowler model (solid line).}
\end{center}
\end{figure}
In general, fragmentation functions $D_{Q\rightarrow M}(z,\mu^2)$ depend on both $z$ and factorization scale $\mu$.
The scale $\mu$  is arbitrary, but in  a high energy process where a jet is produced with transverse momentum $ k_\bot$,
large logarithms of $k_\bot/\mu$ in the parton cross section $ C_i(z, \alpha_s)$ (\ref{fac})
can be avoided by choosing  $\mu$  on the order of $ k_\bot$.
The function (\ref{last}) should be regarded as a model for heavy quark FF at a initial scale $\mu$ of 
order $m_Q$. Here we set the initial scale to $\mu_0=2 m_Q$.
For values of $\mu$ much larger than $\mu_0$, the obtained FF should be evolved from the
scale $\mu_0=2 m_Q$ to the scale $\mu$ using the Altarelli-Parisi equation,
\begin{eqnarray}\label{dgl}
\frac{d}{d\ln \mu^2}D_{Q\rightarrow M}(z,\mu^2)= \int_z^1 \frac{dy}{y}P_{Q\rightarrow Q}(\frac{z}{y},\mu)
D_{Q\rightarrow M}(y,\mu^2),\nonumber\\
\end{eqnarray}
where $P_{Q\rightarrow Q}$ is the appropriate splitting function,
\begin{eqnarray}
P_{Q\rightarrow Q}(x,\mu)=\frac{2\alpha_s(\mu)}{3\pi}\bigg(\frac{1+x^2}{1-x}\bigg)_+.
\end{eqnarray}
As an example we consider the fragmentation  of $c$-quark into the $D^+$-meson
with the constituent quark structure $|D^+> =|c\bar d>$, considering $m_Q=m_c=1.5$ GeV as
was used in \cite{Kneesch:2007ey}, $m_q=m_d=3$ MeV and $\left\langle k_\bot^2 \right\rangle=m_c^2$ in (\ref{last}).
 In Fig.~\ref{comparision} the behavior of $D^+$ FF at the starting scale $\mu_0=2 m_c$ is shown.
 Using the non-perturbative FF  parameters from the second row of Table \ref{fit} 
 and by evolving the FF to the scale $\mu=2 m_c$ 
 we also compare our result with  the Bowler model  as a well-known  phenomenological model \cite{Bowler}. 
 Since to obtain the constant $N$  (\ref{last}) we have used the normalization condition,
 then to compare our result with the Bowler model, our theoretical result should be multiplied by the $c\rightarrow D^+$
 branching fraction $B(2 m_c)=0.235$ \cite{Kneesch:2007ey}, which is defined as $B(\mu)=\int_{z_{cut}}^1dz D(z,\mu^2)$
 where the cut $z_{cut}$ excludes the $z$ range in which  the result is not valid.
As Fig.~\ref{comparision} shows our result is in reliable consistency  with the phenomenological model. 
Note that the function in the pQCD approach is determined in leading order whereas the Bowler function
is extracted in NLO. Therefore, we may also
 think of other effects, such as gluon radiation and secondary fragmentation and so on, which can make
 a better agreement with the phenomenological model. In spite of the uncertainties  mentioned, there is another theoretical
 uncertainty due to the freedom in the choice of scaling variable $z$ which will be discussed in next section.\\
 The $z$ dependences of the FFs  are not yet calculable at each desired scale. However, once
they are given at some initial fragmentation scale $\mu_0$, their $\mu_f$ evolution is specified by
the  evolution equations (\ref{dgl}). Therefore, having the initial FF (\ref{last}), $D_{c\rightarrow D^+}(z,\mu)$
at larger values of $\mu$ can be obtained  by solving DGLAP equations. To illustrate the effects of evolution, we evolve the FF at the
energy scales $\mu=10.52$ GeV and $\mu=m_Z=91.2$ GeV. These results are shown in Fig.~\ref{dglap}.
Since, in the  measurement and the analysis of the data performed by the Belle and the CLEO Collaborations \cite{Seuster:2005tr}, 
the center of mass energy has been set to
$\sqrt{s}=10.52$ GeV, which is much close to the production
threshold of $D$ mesons, we chose this value.
The evolution causes the FF to decrease
at large $z$ and to diverge at $z=0$.\\
Besides the $c\rightarrow D^+$ FF itself, also its first moment is also of phenomenological interest and subject
to experimental determination. It  corresponds to the average  fraction of energy that the $D^+$  meson receives from
the $c$ quark, i.e.
\begin{eqnarray}\label{ave}
\left\langle z \right\rangle (\mu)=\frac{1}{B(\mu)}\int_{z_{cut}}^1 dz z D(z,\mu^2),
\end{eqnarray}
where  $z_{cut}=0.1$.
As is seen from Figs.~\ref{dzero} and \ref{dplus}, there are no experimental data at $z<0.1$.
It is interesting to compare
our result obtained for  the average energy fraction $\left\langle z \right\rangle (m_c)=0.73$  with the values quoted by
BELLE, CLEO, ALEPH and OPAL which are listed  in \cite{Kneesch:2007ey}. There is  good consistency between our result
and the experimental results, however one must keep in mind that experimental results naturally include
all orders and also contributions from gluon and light-quark fragmentation, while ours are evaluated from the $c\rightarrow D^+$
FF at LO via (\ref{ave}).\\
In the remainder of this section, we also compare $z$ distributions of  $D^+$ and $ D^0$ mesons from BELLE and CLEO \cite{Corcella:2007tg} with
our theoretical result. These are shown in Figs.~\ref{dzero} and \ref{dplus}.

\section{Hadron Mass Effects on FFs}
\label{sec:three}

In this section we find it instructive to concentrate  on the massive kinematics of fragmentation,
a topic with a very little attention paid to in the literature. Therefore, we show how to incorporate the
effects of the hadron mass into the fragmentation function using a specific choice of scaling variable.\\
The FF depends on the fragmentation parameter $z$ (\ref{par2}) and
there is some freedom in defining this  parameter in the presence of hadron mass.
In general case the four-momenta of the produced hadron $M$ and the mother  parton $Q$ are related as $p_M=zp_Q$
that in  the  case of massless parton and  hadron
a custom choice is $E_M=z E_Q$ (see (\ref{par3})) in which
the scaling variable $z$ takes the values $0\leq z\leq 1$, i.e. to retain just one of the four equations $p_M=zp_Q$.
This simple relation is not suitable  with the finite quark and/or hadron masses
and needs to be generalized when a heavy quark and/or hadron is considered.
In order to evaluate the theoretical uncertainty due to the freedom in the choice of fragmentation parameter
in the presence of heavy quark and  meson we use an  approach  given in Ref.~\cite{Albino:2005gd}, where
authors took into account the finite mass corrections on the inclusive
hadron production in $e^+e^-$ and hadron-hadron reactions.
For this purpose it is  helpful to work in  light-cone coordinates, in which any four-vector $V$ is written
in the form $V=(V^+, V^-, \vec{V_T})$ where  $V^\pm=(V_0\pm V_3)/\sqrt{2}$ and $\vec{V_T}=(V_1, V_2)$.
Considering (\ref{kinematic}) the momentum of the initial heavy quark takes the form
\begin{eqnarray}
p^\prime=[\frac{p_0^\prime + p_L^\prime}{\sqrt{2}}, \frac{p_0^\prime - p_L^\prime}{\sqrt{2}}, \vec{k_\bot}],
\end{eqnarray}
and for the  massless meson for which $\bar{p}_0=\bar{p}_L$ the four-momentum is expressed as
\begin{eqnarray}\label{olddifenition}
\bar{p}=[\sqrt{2} \bar{p_0}, 0, \vec{0}].
\end{eqnarray}
In the presence of meson and/or quark  masses, the light-cone scaling variable
$\zeta=\bar{p}^+/p^{\prime+}$ seems more convenient than
the fragmentation parameter $z=\bar{p}_0/p_0^\prime$ (\ref{par3}). However,  in
the absence of meson and quark masses, the two variables are identical.
Therefore, to study the effects of hadron mass on FF we apply
the parameter $\zeta$ which  is invariant with
respect to boosts along the three-axis. This axis  is considered as
 the  flight direction of outgoing meson.
Taking mass $M$ for the meson so that $\bar{p}^2=M^2$, the four-momenta
of the meson in the light-cone coordinates reads
\begin{eqnarray}\label{newdifenition}
\bar{p}=[\zeta p^{\prime +}, \frac{M^2}{2 \zeta p^{\prime +}}, \vec{0}].
\end{eqnarray}
Comparison of (\ref{olddifenition}) with (\ref{newdifenition}) shows  that the hadron mass effect is
 imposed  by introducing a non-zero minus component into the hadron's
momentum. From this result we obtain immediately the relation between the two scaling variables
in the presence of hadron mass as
\begin{eqnarray}\label{new}
z=\zeta (1+\frac{M^2}{(2p_0^\prime \zeta)^2 }).
\end{eqnarray}
\begin{figure}
\begin{center}
\includegraphics[width=1.0\linewidth,bb=88 610 322 769]{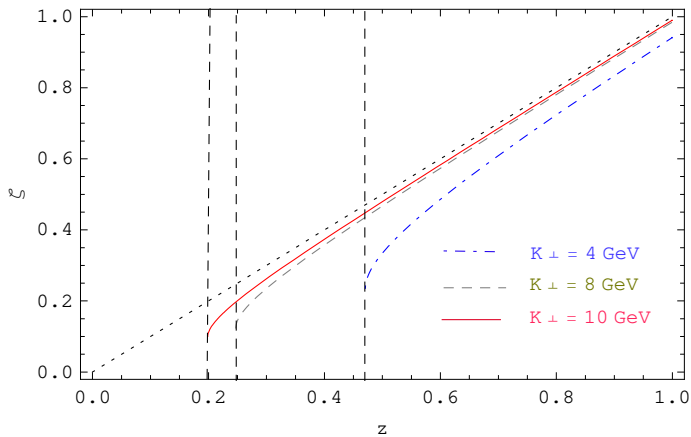}
\caption{\label{variable}%
The variations of $\zeta$  as a function of $z$ for  $k_\bot=4$ GeV,  $k_\bot=8$ GeV
and  $k_\bot=10$ GeV. Thresholds are also shown.}
\end{center}
\end{figure}
\begin{figure}
\begin{center}
\includegraphics[width=1.0\linewidth,bb=88 610 322 769]{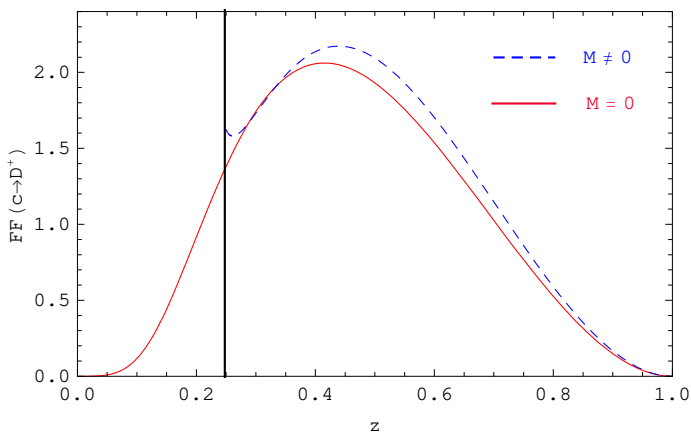}
\caption{\label{mass}%
$c\rightarrow D^+$ FF as a function of $z$ when $M=0$ (solid line)
and $M\neq 0$ (dashed line) taking $k_\bot=8$ GeV. Threshold at $z=0.25$ is also shown.}
\end{center}
\end{figure}
Note that these two variables are  equal when $M\rightarrow 0$.
Considering the four-momenta from (\ref{kinematic})
and at the fragmentation process with the sufficiently large transverse momentum, one can write $m_Q^2\approx {p_0^\prime}^2-k_\bot^2$.
In Fig.~\ref{variable}, taking $m_Q=m_c=1.5$ GeV and $M=m_{D^+}=1.87$ GeV  \cite{Nakamura:2010zzi}
the variations of new scaling variable $\zeta$ as a function of  $z$  is shown for different values of $k_\bot$.
As it is seen the effect of meson mass is considerable when  the transverse momentum  $k_\bot$
decreases and this effect also creates a threshold for the FFs.\\
Now, to obtain the improved FF we go back to the definition of
fragmentation (\ref{first}), i.e. $D_{Q\rightarrow M}(z)=1/\sigma \cdot d\sigma(z)/dz$.
As a generalization of the massless hadron case, we assume that the cross section which we have been
calculating is $d\sigma(\zeta)/d\zeta$ which is related to the measured observable $d\sigma(z)/dz$ via
\begin{eqnarray}
\frac{d\sigma(z)}{dz}=\frac{d\zeta}{dz}\frac{d\sigma(\zeta(z))}{d\zeta},
\end{eqnarray}
that is simplified as
\begin{eqnarray}
\frac{d\sigma(z)}{dz}=\frac{1}{1-(\frac{M}{2p_0^\prime\zeta})^2}\frac{d\sigma(\zeta(z))}{d\zeta}.
\end{eqnarray}
Here, using (\ref{new}) our new scaling variable is expressed as
\begin{eqnarray}
\zeta=\frac{z}{2}\big\{1+(1-\frac{M^2}{z^2(m_Q^2+k_\bot^2)})^{\frac{1}{2}}\big\},
\end{eqnarray}
and now  the observable quantity $D_{Q\rightarrow M}(z)$,  reads
\begin{eqnarray}\label{improved}
D_{Q\rightarrow M}(z,\mu)=\frac{1}{1-\frac{M^2}{4\zeta^2(m_Q^2+k_\bot^2)}}D_{Q\rightarrow M}(\zeta(z), \mu).
\end{eqnarray}
Note that  $D_{Q\rightarrow M}(\zeta, \mu)$ is the FF obtained in (\ref{last}) by substituting $z\rightarrow \zeta$ and
the kinematically allowed $z$ ranges are now  $M/\sqrt{m_Q^2+k_\bot^2}<z\leq 1$.\\
In Fig.~\ref{mass}  the behavior of $D^+$ fragmentation function is shown for the massless
and  massive meson considering $m_{D^+}=1869.62$ MeV. As is shown the effect of meson
 mass is increasing the size of FF at large values of $z$
and the peak position is shifted towards higher values of $z$ and it also creates
a threshold at $z=0.25$.

\section{Conclusion}
\label{sec:four}

We  studied the heavy quark FFs in the current approaches. Using the perturbative
QCD scheme we presented an analytical form for the FFs to produce S-wave heavy mesons to leading order
 in $\alpha_s$ which agrees with most kinematical and dynamical expectations. Our result
describes not only the $z$ dependence of the fragmentation probabilities, but also their dependence
on the transverse momentum of the meson relative to the  produced jet.
The perturbative QCD FF was  compared  with a well-known phenomenological   
model for the heavy quark fragmentation in the literature.
Specifically, we  compared the FFs for $D^0$ and $D^+$ mesons with available $e^-e^+$
 annihilation data, from CLEO and BELLE \cite{Corcella:2007tg} and we found good agreement.
We also investigated, for the first time,  finite meson mass corrections on the pQCD FFs
and their theoretical uncertainty due to the freedom
in the choice of the scaling variable.
The advent of precise data from $D$ factories motivates the incorporation of hadron mass
effect, which are then likely to be no longer negligible into the formalism.

\end{document}